\documentclass[prd,showkeys,floatfix,twocolumn,amsmath,amssymb,floatfix]{revtex4}
\usepackage{amsmath}
\usepackage{graphicx}
\usepackage{float}
\usepackage{epstopdf}
\usepackage{multirow}
\usepackage{subfigure}
\usepackage{color}
\usepackage{dashrule}
\usepackage[colorlinks,citecolor=blue,anchorcolor=red,menucolor=red,linkcolor=red,filecolor=red,runcolor=red,urlcolor=blue,frenchlinks=red]{hyperref}
\usepackage{orcidlink}

\allowdisplaybreaks[3]

\newcounter{cnt}
\makeatletter

\let\oldhypertarget\hypertarget
\renewcommand{\hypertarget}[2]{%
  \oldhypertarget{#1}{#2}%
    \protected@write\@mainaux{}{%
        \string\expandafter\string\gdef
          \string\csname\string\detokenize{#1}\string\endcsname{#2}%
    }%
  }
\newcommand{\myhyperlink}[1]{%
  \hyperlink{#1}{\csname #1\endcsname}%
  }
\makeatother

\newcommand{\cref}[1]{\myhyperlink{#1}}

\renewcommand{\arraystretch}{1.8}

\begin{document}


\title{Heavy flavored hydrogen molecule systems}

\author{Hui-Min Yang\orcidlink{0000-0001-9673-5623}$^1$}
\email{hmyang@pku.edu.cn}
\author{Yao Ma\orcidlink{0000-0002-5868-1166}$^1$}
\email{yaoma@pku.edu.cn}
\author{Shi-Lin Zhu\orcidlink{0000-0002-4055-6906}$^1$}
\email{zhusl@pku.edu.cn}

\affiliation{
$^1$School of Physics and Center of High Energy Physics, Peking University, Beijing 100871, China \\
}

\begin{abstract}
This study provides a comprehensive analysis of $S$-wave exotic hydrogen-like three-body systems ($pp\mu^-$, $pp\tau^-$, $\mu^-\mu^-p$, $\tau^-\tau^-p$, $p\mu^-\tau^-$) with spin-parity $J^P = 1/2^+$ and $3/2^+$, and four-body systems ($pp\mu^-\mu^-$, $pp\tau^-\tau^-$) with $J^P = 0^+$, $1^+$, and $2^+$. The wave functions are constructed using complete spin configurations and spatial components with various Jacobi coordinate structures. We use complex scaling and Gaussian expansion methods to solve the complex-scaled Schr\"{o}dinger equation and obtain possible bound and quasi-bound states. The resulting binding energies range from $-33.8$~keV to $-340$~eV, with the $p\mu^-\tau^-$ system being the only case where no bound state is found. Notably, we present the first theoretical estimation of the bound-state energy levels of $pp\mu^-\mu^-$ and $pp\tau^-\tau^-$, which is of significant importance for understanding exotic few-body Coulomb systems. We further analyze spin configurations and root-mean-square radii to elucidate the spatial structure of these bound and quasi-bound states. Our results reveal that $K$-type spatial configurations play a crucial role in accurately describing bound and quasi-bound states in the hydrogen-molecule-like systems $pp\mu^-\mu^-$ and $pp\tau^-\tau^-$. Incorporating $K$-type configurations significantly alters the mass spectra of these states. Future muon colliders and muon facilities may offer promising platforms for the possible copious production of such heavy flavored hydrogen molecules and molecular ions. For instance, scattering processes such as $2\mu^- + \mathrm{H_2} \to \mathrm{H_{2\mu}} + 2e^-$, $\mu^- + \mathrm{H_2} \to \mathrm{H_{\mu e}} + e^-$, and $\mu^- + \mathrm{H_2^+} \to \mathrm{H_{2\mu}^+} + e^-$ could be utilized, facilitating detailed studies of intriguing states such as $\mathrm{H_{2\mu}}$, $\mathrm{H_{\mu e}}$, and $\mathrm{H_{2\mu}^+}$.
\end{abstract}
\maketitle
\pagenumbering{arabic}
%
%
%
\section{Introduction}\label{sec:intro}
%
For a long time, a central objective of atomic, molecular  and nuclear physics has been to uncover the microscopic structure of atoms, ions, and molecules, as well as the fundamental laws governing the interactions among their constituent particles (i.e., nucleons and electrons)~\cite{Ramos:2025tge,Li:2025gnw,Ma:2024svb,Anderson:2022jhq,Makii:2019tkb,Cheng:2024mxh,Carlson:2014vla,Navratil:2016ycn,Dobaczewski:1997cu,Dobaczewski:2007jc,Forssen:2012yn,Balantekin:2014opa}. In this context, electromagnetic interactions characterized by the Coulomb force, spin-spin, and spin-orbit interaction terms, have the potential to form bound states and quasi-bound states~\cite{Wang:2017qij,Schwierz:2007ve,Woods:1954zz,Thompson:1977zz,Dudek:1982zz}. Conventional hydrogenic systems, such as $H^-$, $H_2^+$, $D_2^+$, $T_2^+$, $H_2$, have been extensively studied~\cite{moiseyev1979autoionizing,Singh:2024abj,OMalley:2016ugf,Seeley:2012zpt,PhysRevApplied.11.044092,Urey:1932gik,Jentschura:2011mdl}. In contrast, exotic hydrogen-like systems, where one or more electrons are replaced by heavier leptons (muons or taus), have received comparatively less attention~\cite{moiseyev1979autoionizing,liverts2013three,Rojas:2021kkb}. These exotic systems provide a unique platform for testing fundamental physics. 

The simplest muonic atoms $p\mu^-$ has driven precision tests of QED. The CREMA collaboration’s measurements of the muonic hydrogen Lamb shift gave rise to the proton radius puzzle and reported a $7\sigma$ discrepancy between muonic and electronic determinations, potentially indicating limitations in the Standard Model or nuclear physics models~\cite{pohl2013muonic}. 
For three-body systems, theoretical interest surged following Vesman’s 1967 prediction of the metastable $pp\mu^-$ molecular ion, a system where the muon catalyzes nuclear fusion~\cite{vesman1967pis}. Subsequent studies employing variational methods and adiabatic approximations refined the understanding of its energy levels, decay rates, and fusion dynamics~\cite{pohl2010size,pohl2016laser,pachucki1996theory,pohl2013muonic,jentschura2005calculation}. For systems with two heavy leptons, theoretical frameworks have been developed that integrate few-body quantum mechanics with QED. These studies predict enhanced sensitivity to vacuum polarization, self-energy corrections, and nuclear finite-size effects~\cite{pachucki1996theory,jentschura2005calculation}. The $\mu^-\mu^-p$ atom, for instance, exhibits a ``supercharged" nucleus where the muons orbit approximately 200 times closer to the nucleus than electrons, amplifying QED shifts by orders of magnitude. Tauonic systems such as $pp\tau^-$, $\tau^-\tau^-p$ remain largely unexplored experimentally due to the tau’s short lifetime ($\sim 10^{-13}$ s). Nevertheless, theoretical models suggest these systems could serve as sensitive probes of LFU violation and anomalous magnetic moments.

Another important physical problem is the study of bound state in these exotic systems. Bound states such as $pp\mu^-$, $pp\tau^-$, $\mu^-\mu^-p$, $\tau^-\tau^-p$, $pp\mu^-\mu^-$, and $pp\tau^-\tau^-$ enable precision studies of quantum electrodynamics (QED), electroweak interactions, and nuclear structure under extreme conditions, which are characterized by high reduced masses, compact wavefunctions, and significant relativistic and radiative corrections. These systems offer exceptional sensitivity to bound-state QED effects, tests of lepton flavor universality (LFU), and nuclear charge distributions, thus complementing conventional atomic spectroscopy. Over the past decades, a wide range of theoretical techniques have been developed to investigate their structure and dynamics. Among them, several models and computational methods have been employed to explore various physical properties. For instance, in the hydrogen molecule, the Born-Oppenheimer (BO) approximation~\cite{born1985quantentheorie,Hagedorn:1980fp,Klein:1993gi,Mehta:2014tfa,Zettili:1986nz,Stegeby:2012xj} assumes a separation of electronic and nuclear motion, effectively reducing a high-dimensional system into two decoupled subsystems of lower dimension. Within this framework, the total energy of the hydrogen molecule is decomposed into contributions from the electronic, vibrational, and rotational excitation energies~\cite{flugge2012practical,moiseyev1979autoionizing}. The discovery of strong electron-electron correlation effects in doubly excited resonant states of helium stimulated the development of group-theoretical and adiabatic quantum approximations to understand these correlations~\cite{Tanner:2000zz}. The bound state spectra of two-electron atoms could be calculated efficiently with the help of the Hartree-Fock self-consistent-field method after the advent of modern quantum theory. Additionally, the correlation function hyperspherical harmonic method has been explored in Ref.~\cite{PhysRevA.42.6324}. 

While the majority of theoretical investigations have focused primarily on bound-state problems, quasi-bound states have received comparatively less attention.
A number of studies have employed the complex scaling method (CSM) to the study of quasi-bound states in atomic systems~\cite{simon1971convergence,simon1973resonances,Aguilar:1971ve,Balslev:1971vb,Aoyama:2006hrz}. These studies include the calculation of autoionizing states of $H_2^-$ and $H_2$ ion~\cite{doolen1975procedure,ho1977autoionisation,burgers2000doubly}, as well as autoionization states of the positronium negative ion and some $\mu$-meson ions~\cite{ho1979autoionization,ho1983method,nagashima2014experiments,reinhardt1982complex,mills1981observation,junker1982recent,ho1981complex,bhatia1983new,Ma:2025rvj}. The bound and quasi-bound states of exotic molecules such as $pp\mu^-$ have also been reported~\cite{shimamura1989series,liverts2013three,PhysRevA.59.4270,frolov1993algebraic,ho1979autoionisation,ho1981complex,kilic2004coulombic}.

To extend these investigations, we aim to systematically explore resonant states in a broader variety of three- and four-body exotic hydrogen-like systems. Notably, studies of the doubly heavy state $T_{cc}(3875)$~\cite{LHCb:2021vvq,LHCb:2021auc} offer valuable methodological insights for exploring four-body hydrogen-like molecular systems such as $pp\mu^-\mu^-$ and $pp\tau^-\tau^-$. The $T_{cc}(3875)$, composed of two heavy quarks and two light quarks, may serve as a shallow bound state with quantum numbers $I(J^P)=0(1^+)$. In our calculation, the $pp\mu^-\mu^-$ and $pp\tau^-\tau^-$ systems, which similarly contain two heavy and two light particles, are found to form bound states only with $J^P=0^+$. This discrepancy is likely attributable to the complex chromoelectric and chromomagnetic interactions in the QCD-based systems, in contrast to the purely Coulombic interactions governing QED systems.

In this work, we conduct a systematic study of $S$-wave hydrogen-like systems, including three-body systems ($pp\mu^-$, $pp\tau^-$, $\mu^-\mu^-p$, $\tau^-\tau^-p$, $p\mu^-\tau^-$) with spin-parity $J^P=1/2^+$, $3/2^+$, and four-body systems ($pp\mu^-\mu^-$, $pp\tau^-\tau^-$) with $J^P=0^+$, $1^+$ and $2^+$. We construct the wave functions of these systems using a complete set of spin configurations and spatial wave functions under different Jacobi coordinate configurations. To predict possible bound and quasi-bound states, we employ the CSM~\cite{Aguilar:1971ve,Balslev:1971vb,Aoyama:2006hrz} and Gaussian expansion method~\cite{Hiyama:2003cu} (GEM) to solve the complex-scaled Schr\"{o}dinger equation. Spectra, spin configurations and rms radii are computed to investigate the properties of exotic states, offering deeper insight into the internal dynamics of hydrogen-like systems.

This paper is organized as follows. In Sec.~\ref{sec:formulation}, we briefly introduce our theoretical framework, including GEM and CSM. We construct the wave functions of $S$-wave three- and four-body hydrogen-like systems. Additionally, we use the rms radii to describe the spatial properties of bound and quasi-bound states. In Sec.~\ref{sec:results}-\ref{sec:results-1}, we apply GEM and CSM to calculate the complex eigenenergies of these exotic hydrogen-like systems and analyze the properties of the resulting states. Finally, we present our conclusions in Sec.~\ref{sec:summary}.

%
\section{Formulation}
\label{sec:formulation}
%
\subsection{Hamiltonian}
In QED $n-$body system the nonrelativisitic Hamiltonian reads
\begin{eqnarray}
H&=& \sum_{i=1}^n({p_i^2\over 2m_i}+m_i)-T_{CM}+\sum_{i<j=1}^n V_{ij}\, ,
\end{eqnarray}
where $m_i$ and $p_i$ represent the mass and momentum of the $i$-th particle, respectively. $T_{CM}$ denotes the center-of-mass kinematic energy, which has been subtracted in our calculation. $V_{ij}$ represents the interaction between the $i$-th and $j$-th particles. In this work, we apply the Coulomb potential to study QED systems with two protons and one or two leptons.
\begin{eqnarray}
 V_{ij}&=& \frac{\alpha Q_i Q_j}{r_{ij}}\, ,
\end{eqnarray}
The $\alpha$ is the fine-structure constant. The $Q_i$ denotes the charge of the $i-$th particle. $r_{ij}$ refers to the relative position between the $i$-th and $j$-th particles. We calculate the masses and root-mean-square radii of hydrogen-like atoms and collect them in Table~\ref{tab:mass}.
\begin{table}
\begin{center}
\renewcommand{\arraystretch}{1.5}
\caption{The binding energy $\Delta E$ and rms radii $r^{\rm rms}$ of $pl^-$ systems. }
\begin{tabular}{ c  c  c  c}
\hline\hline
$J^{P}$& ~~~~System~~~~ & ~~~$\Delta E$~~~ &~~~$r^{\rm rms}$
\\ \hline
$0^+/1^+$&$\rm H(1S)$&$-13.6~\rm eV$&$0.09~\rm nm$
\\ \cline{2-4}
&$\rm H(2S)$&$-3.4~\rm eV$&$0.34~\rm nm$
\\ \cline{2-4}
&$\rm H(3S)$&$-1.5~\rm eV$&$0.76~\rm nm$
\\ \cline{2-4}
&$p\mu^-(1S)$&$-2.53~\rm keV$&$0.49~\rm pm$
\\ \cline{2-4}
&$p\mu^-(2S)$&$-0.63~\rm keV$&$1.85~\rm pm$
\\ \cline{2-4}
&$p\mu^-(3S)$&$-0.28~\rm keV$&$4.10~\rm pm$
\\ \cline{2-4}
&$p\tau^-(1S)$&$-16.35~\rm keV$&$0.08~\rm pm$
\\ \cline{2-4}
&$p\tau^-(2S)$&$-4.09~\rm keV$&$0.29~\rm pm$
\\ \cline{2-4}
&$p\tau^-(3S)$&$-1.82~\rm keV$&$0.63~\rm pm$
\\ \hline\hline
\end{tabular}
\label{tab:mass}
\end{center}
\end{table}

\subsection{Wave function}

The wave function $\Psi$ of three-body systems $pp l^-$, $l^-l^{(')-} p$ and four-body systems $ppl^-l^-$ can be expanded by the direct product of the spin wave function $\chi_s$ and the spatial wave function $\phi$
\begin{eqnarray}
\Psi=\mathcal{A}(\chi_s\otimes \phi)\, ,
\end{eqnarray}
where $\mathcal{A}$ is the anti-symmetrization operator for identical particles. The particles in hydrogen-like systems are labeled sequentially as $1$, $2$, $3$ (and $4$). For the $ppl^-$ and $l^-l^{(')-}p$ systems, the operator $\mathcal{A}=1-P_{12}$; for $ppl^-l^-$, $\mathcal{A}=(1-P_{12})(1-P_{34})$, where $P_{ij}$ represents the permutation operator that exchanges the $i$-th and $j$-th particles.
 
\begin{figure}[H]
\begin{center}
\subfigure[]{
\scalebox{0.5}{\includegraphics{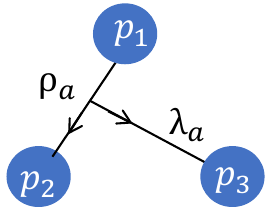}}}~~~~
\subfigure[]{
\scalebox{0.5}{\includegraphics{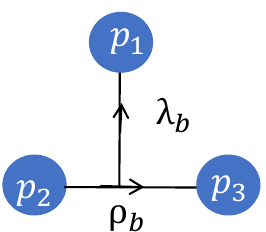}}}~~~~
\subfigure[]{
\scalebox{0.5}{\includegraphics{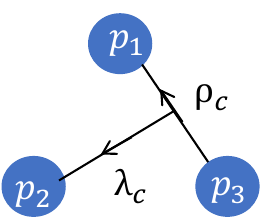}}}
\caption{The Jacobi coordinates for three types of spatial configurations in three-body QED systems.}
\label{fig:Jacobi-three}
\end{center}
\end{figure}

\begin{figure}[H]
\begin{center}
\subfigure[]{
\scalebox{0.6}{\includegraphics{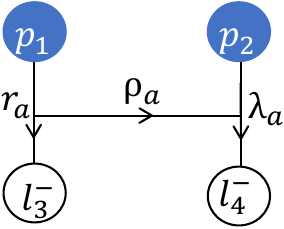}}}~~~~
\subfigure[]{
\scalebox{0.6}{\includegraphics{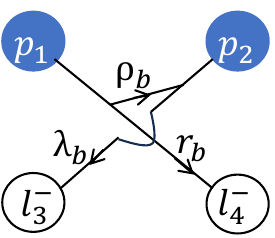}}}
\\
\subfigure[]{
\scalebox{0.6}{\includegraphics{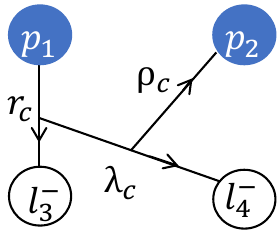}}}~~~~
\subfigure[]{
\scalebox{0.6}{\includegraphics{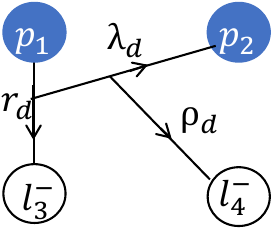}}}
\\
\subfigure[]{
\scalebox{0.6}{\includegraphics{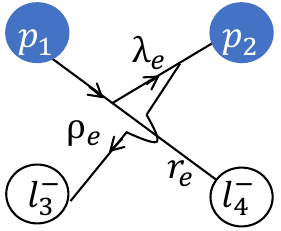}}}~~~~
\subfigure[]{
\scalebox{0.6}{\includegraphics{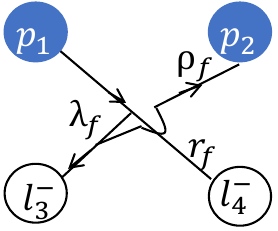}}}
\\
\subfigure[]{
\scalebox{0.6}{\includegraphics{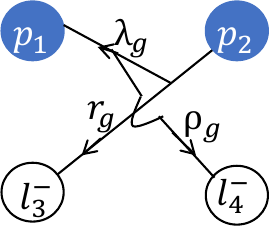}}}~~~~
\subfigure[]{
\scalebox{0.6}{\includegraphics{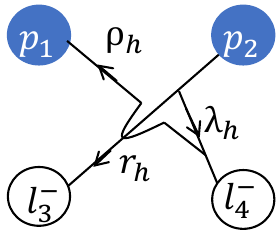}}}
\\
\subfigure[]{
\scalebox{0.6}{\includegraphics{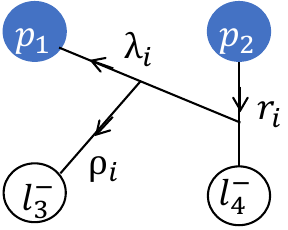}}}~~~~
\subfigure[]{
\scalebox{0.6}{\includegraphics{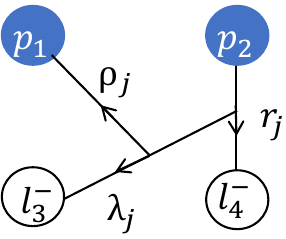}}}
\caption{The Jacobi coordinates for eight types of spatial configurations in four-body hydrogen-like systems.}
\label{fig:Jacobi-four}
\end{center}
\end{figure}

We construct various types of spatial configurations for three-body and four-body QED systems, focusing exclusively on $S$-wave case. For three-body systems, we consider three distinct $S$-wave Jacobi configurations. These include the $T$-type structures, where two particles form a cluster and the third particle couples to it. The corresponding spatial structures are described using Jacobi coordinates $\rho_\alpha$ and $\lambda_\alpha$, as illustrated in Fig.~\ref{fig:Jacobi-three}. For four-body QED systems, we similarly construct $S$-wave spatial configurations using different Jacobi coordinate arrangements, including dihydrogen-like atomic structures and $K$-type configurations. These configurations are described by Jacobi coordinates $r_\alpha$, $\lambda_\alpha$, and $\rho_\alpha$, as shown in Fig.~\ref{fig:Jacobi-four}. Diparticle–diparticle configurations in which each cluster contains two particles with identical charges are excluded due to the repulsive electromagnetic interaction, which causes the particles to move apart. In contract, four-body QCD systems require such a configuration, namely the diquark-antidiquark structure, as the complex chromoelectric and chromomagnetic potentials may lead to attractive interactions between two quarks (antiquarks). Additionally, $K$-type configurations are not only appropriate but sometimes essential in four-body QED systems or nuclear physics, where long-range electromagnetic or residual strong interactions arise between a single particle and the remaining three particles. Therefore, $K$-type configurations are included in this work. To mitigate the computational complexity in the $K$-type spatial configurations, we restrict the $K$-type wave functions to two dominant Jacobi coordinate sets: $[[p\l^-]p]\l^-$ and $[[p\l^-]\l^-]p$. Notably, for system containing identical particles, one spatial configuration may transform into another through particle exchange symmetry. In this work, the spatial wave function adopts the superposition of Gaussian bases as
\begin{eqnarray}
\label{eq:gem}
\phi_{nml}&=&\sqrt{2^{l+5/2}\over \Gamma(l+3/2)r_n^3}({r\over r_n})^l e^{-{r^2\over r_n^2}}Y_{lm}(\hat{r})\, ,
\\ \nonumber r_n&=& r_0 a^{n-1}\, .
\end{eqnarray} 

In principle, one can select a specific Jacobi coordinate and construct a complete set of basis functions with total orbital angular momentum $l$. However, treatment of angular momentum in GEM remains technically challenging, despite the availability of some established strategies~\cite{Hiyama:2003cu}. In this work, we only restrict the spatial wave functions to $l=0$ in Eq.~\ref{eq:gem}, while incorporating different Jacobi coordinate configurations. In this way, orbital excitations can effectively emerge when one Jacobi configuration is transformed into another.

For spin wave functions with total angular momentum $J$, we choose a complete set of basis
\begin{itemize}
\item three-body QED system
\begin{eqnarray}
\chi_1^{J=1/2}&=&[(p_1 p_2)_{0}l^-_{3}]_{1/2}\, ,
\\  
\chi_2^{J=1/2}&=&[(p_1 p_2)_{1}l^-_{3}]_{1/2}\, ,
\\ 
\chi_3^{J=3/2}&=&[(p_1 p_2)_{1}l^-_{3}]_{3/2}\, ,
\\ 
\chi_4^{J=1/2}&=&[(l^-_1 l^-_2)_{0}p_{3}]_{1/2}\, ,
\\  
\chi_5^{J=1/2}&=&[(l^-_1 l^-_2)_{1}p_{3}]_{1/2}\, ,
\\ 
\chi_6^{J=3/2}&=&[(l^-_1 l^-_2)_{1}p_{3}]_{3/2}\, .
\end{eqnarray}

\item four-body QED system
\begin{eqnarray}
\chi_7^{J=0}&=&[(p_1 p_2)_{0}(l^-_3 l^-_4)_{0}]_{0}\, ,
\\
\chi_8^{J=0}&=&[(p_1 p_2)_{1}(l^-_3 l^-_4)_{1}]_{0}\, , 
\\  
\chi_9^{J=1}&=&[(p_1 p_2)_{0}(l^-_3 l^-_4)_{1}]_{1}\, ,
\\ 
\chi_{10}^{J=1}&=&[(p_1 p_2)_{1}(l^-_3 l^-_4)_{0}]_{1}\, ,
\\ 
\chi_{11}^{J=1}&=&[(p_1 p_2)_{1}(l^-_3 l^-_4)_{1}]_{1}\, ,
\\ 
\chi_{12}^{J=2}&=&[(p_1 p_2)_{1}(l^-_3 l^-_4)_{1}]_{2}\, .
\end{eqnarray}
\end{itemize}

\subsection{Complex scaling method}

The complex scaling method (CSM), originally proposed by Aguilar, Balslev, and Combes, has been widely used to directly extract resonance energies and decay widths in atomic and molecular physics~\cite{Aguilar:1971ve,Balslev:1971vb,Aoyama:2006hrz}. In recent studies~\cite{Wang:2022yes,Chen:2023syh,Ma:2024vsi,Wu:2024euj,Wu:2024hrv,Wu:2024zbx,Ma:2025rvj,Yang:2025wqo,Wu:2024ocq}, CSM has also been applied to identify possible bound and resonant states in few-body QED and QCD systems.

Following these works, we perform a systematic analysis of $S$-wave QED three-body systems $pp l^-$, $l^-l^{(')-} p$ with quantum number $J^P=1/2^+,3/2^+$, and four-body system $pp l^-l^-$ with quantum number $J^P=0^+,1^+,2^+$. The CSM introduces the transformation $U(\theta)$ for the radial coordinate $r$ and its conjugate momentum $p$
\begin{eqnarray}
\nonumber U(\theta)r=r e^{i\theta}\, ,~~~~U(\theta)p=p e^{-i\theta}
\end{eqnarray}

The Hamiltonian is transformed as
\begin{eqnarray}
\nonumber H(\theta)&=& \sum_{i=1}^n\Big(m_i+{p_i^2e^{-2i\theta}\over 2m_i}\Big)-T_{CM}+\sum_{i\textless j=1}^n V_{ij}(r_{ij}e^{i\theta})
\end{eqnarray}

Then we solve the complex-scaled $n$-body Schr\"{o}dinger equation. The wave function of $S$-wave three-body and four-body systems with total angular momentum $J$ are expressed as
\begin{equation}
\Psi^J(\theta)_{3-body}=\mathcal{A}\sum_{\alpha,n_i,\beta}C_{\alpha,n_i,\beta}(\theta)\chi_\alpha^J\phi_{n_1}(\rho_\beta)\phi_{n_2}(\lambda_\beta)\, ,
\end{equation}
\begin{equation}
\Psi^J(\theta)_{4-body}=\mathcal{A}\sum_{\alpha,n_i,\beta}C_{\alpha,n_i,\beta}(\theta)\chi_\alpha^J\phi_{n_1}(r_\beta)\phi_{n_2}(\lambda_\beta)\phi_{n_3}(\rho_\beta)\, ,
\end{equation}
where $C_{\alpha,n_i,\beta}(\theta)$ denotes the complex expansion coefficient of basis function, which is determined by solving the energy eigenvalue equation. The spatial wave function $\phi_{n_i}(r_\beta)$ is a real-range Gaussian basis function and does not undergo the operator $U(\theta)$ transformation. Instead, the scaling angle $\theta$ is incorporated into the eigenvalue $E(\theta)$, the coefficient $C_{\alpha,n_i,\beta}(\theta)$ and Hamiltonian $H(\theta)$. The complex-scaled Schr\"{o}dinger equation is expressed as
\begin{eqnarray}
H(\theta)\Psi^J_{n-body}(\theta)&=& E(\theta)\Psi^J_{n-body}(\theta)
\end{eqnarray}

According to the ABC theorem~\cite{Aguilar:1971ve,Balslev:1971vb}, eigenenergies of continuum (scattering), bound, and resonant states can be determined by solving the complex-scaled Schr\"{o}dinger equations. In the complex energy plane, continuum states are aligned along rays originating at the threshold energies, rotated clockwise by an angle $\operatorname{Arg}(E)=2\theta$. Bound states appear on the negative real axis. Resonant states locate between the real axis and the continuum lines and correspond to poles ($E_R=M_R-i\Gamma_R/2$) on the second Riemann sheet associated with the respective thresholds. Both the bound and resonant states are independent of the scaling angle $\theta$ and remain stable as it varies.

\subsection{Spatial structure}

In addition to fundamental physical properties such as mass and width, the complex spatial structure is a crucial feature for characterizing the enigmatic nature of hydrogen-like systems. To explore their internal configuration, we investigate the spatial distributions of these systems, with rms radius defined as
\begin{eqnarray}
r_{ij}^{rms,C}&=& \operatorname{Re}\Big[\sqrt{{\langle \Psi(\theta)|r^2_{ij}e^{2i\theta}|\Psi(\theta)\rangle\over \langle\Psi(\theta)|\Psi(\theta)\rangle}}\Big]\, .
\end{eqnarray}

Within the framework of CSM, inner products are defined using the c-product as
\begin{equation}
\langle \phi_n|\phi_m\rangle=\int \phi_n(r) \phi_m (r)d^3 r\, .
\end{equation}
The rms radius of resonant states calculated using the c-product generally possesses both real and imaginary components. As discussed in Ref.~\cite{homma1997matrix}, the real part reflects the expectation value of the spatial distribution between particles, while the imaginary part represents the uncertainty associated with the measurement, assuming the resonance is not excessively broad.

\section{Three-body hydrogen-like systems}
\label{sec:results}

We employ the CSM to calculate the complex eigenenergies of the $S$-wave three-body hydrogen-like systems, including $pp\mu^-$, $pp\tau^-$, $\mu^-\mu^- p$, $\tau^-\tau^- p$, and $p\mu^-\tau^-$, with spin-parity quantum numbers $J^{P}=1/2^+$ and $3/2^+$. The distribution of complex eigenenergies for these systems is shown in Figs.~\ref{fig:ppmu}-\ref{fig:pmutau}, respectively. These figures exhibit several characteristic features. The bound states lie on the negative real axis of energy plane. The continuum states are located along the continuum lines originating at the thresholds, which are rotated clockwise by $2\theta$ from the positive real axis. Additionally, there are resonant states situated between the positive real axis and the continuum lines, corresponding to poles on the second Riemann sheet of the respective thresholds. The bound and resonant states do not shift with the variation of the angle $\theta$, and are circled by black circles. The complex eigenenergies, proportions of spin configurations and rms radii of the bound and resonant states are listed in Tables~\ref{tab:ppl}-\ref{tab:pmutau}.

\subsection{$pp\mu^-$ and $pp\tau^-$}

\begin{figure}[hbtp]
\begin{center}
\scalebox{0.35}{\includegraphics{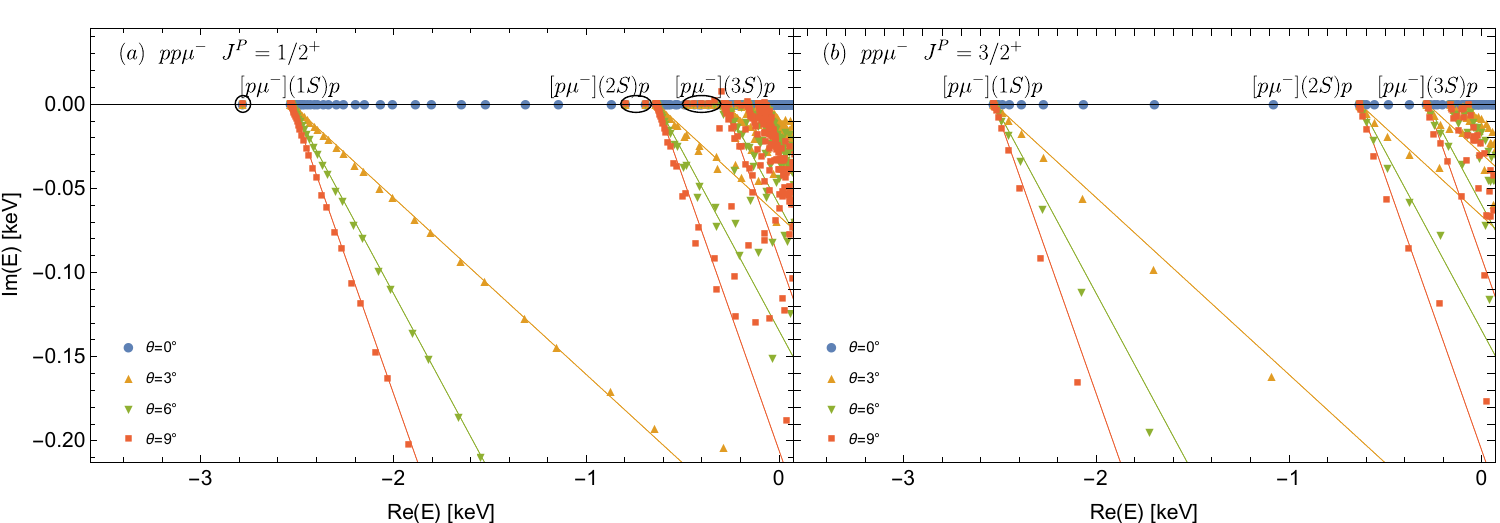}}
\end{center}
\caption{The complex eigenenergies of the $pp\mu^-$ states with varying $\theta$ in the CSM. The solid lines represent the continuum lines rotating along $Arg(E)=-2\theta$. The bound states and resonant states remain stable as the scaling angle $\theta$ changes, which are marked out by the black circles. }
\label{fig:ppmu}
\end{figure}

The $S$-wave $pp\mu^-$ system, regarded as a hydrogen-like molecular ion, has attracted significant attention in nuclear physics. We identify one bound and six resonant states (or quasi-bound) with quantum numbers $J^P=1/2^+$, along with continuum states for $J^P=3/2^+$, respectively. The binding energies $\Delta E$ of bound and quasi-bound states range from $-2.8$~keV to $-0.3$~keV relative to the three-body threshold. The bound state lies below the lowest $[p\mu^-](1S)p$ threshold and the six quasi-bound states are located below the $[p\mu^-](3S)p$ threshold, as shown in Fig.~\ref{fig:ppmu}. Their complex energies, spin configuration, and rms radii are summarized in Table~\ref{tab:ppl}. It is noted that the tiny weak decay width of $\mu^-$ is not involved in the total decay width. To facilitate analysis of spin configurations, we label the three particles in the $pp\mu^-$ system as 1, 2, and 3. The spin structure is denoted as $[s_{12},s_3]_{s}$, where $s_{12}$ represents the total spin resulting from the coupling of $s_1$ and $s_2$, and $s$ denotes the total spin of the system.

\begin{table*}[hbtp]
\begin{center}
\renewcommand{\arraystretch}{1.5}
\caption{The complex eigenenergies $\Delta E-i{\Gamma/2}$, spin components, and rms radii between different particles of the three-body hydrogen-like $pp l^-$ systems. Here, $\Delta E-i{\Gamma/2}$ represents the binding energy relative to the three body threshold. In the third column, “B” indicates bound states, while “R” denotes quasi-bound states. The spin configuration is expressed as $[s_{12},s_3]_{s}$. In the last column, we list the binding energies of the corresponding states extracted from other works for comparison.
}
\begin{tabular*}{\hsize}{@{}@{\extracolsep{\fill}}cccccccc@{}}
\hline\hline
system &$I(J^P)$&~~~$\Delta E-i\Gamma/2(\rm eV) $ ~~~&~~type ~~&~~~spin configuration~~~&~~~$r_{pp}^{\rm{rms}}$(pm)~~~&~~~$r_{pl^-}^{\rm{rms}}$(pm)~~& $\Delta E-i\Gamma/2(\rm eV) $
\\  \hline
$pp\mu^-$&$1(\frac{1}{2}^+)$& $-2780$& B &$[0,{1\over 2}]_{1\over 2}$&$0.90$&$0.71$& $-2782$~\cite{liverts2013three,kilic2004coulombic,PhysRevA.59.4270}
\\ \cline{3-8}
&& $-793$& R &$[0,{1\over 2}]_{1\over 2}$&$2.94$&$1.93$& $-824$~\cite{liverts2013three,kilic2004coulombic}
\\ \cline{3-8}
&& $-691$& R &$[0,{1\over 2}]_{1\over 2}$&$3.95$&$2.59$&$\cdots$
\\ \cline{3-8}
&&$-474$& R &$[0,{1\over 2}]_{1\over 2}$&$4.57$&$3.70$&$\cdots$
\\ \cline{3-8}
&&$-437-0.27i$& R &$[0,{1\over 2}]_{1\over 2}$&$5.17$&$3.79$&$\cdots$
\\ \cline{3-8}
&&$-399-0.32i$& R &$[0,{1\over 2}]_{1\over 2}$&$6.08$&$4.10$&$\cdots$
\\ \cline{3-8}
&&$-363-0.12i$& R &$[0,{1\over 2}]_{1\over 2}$&$7.13$&$4.52$&$\cdots$
\\ \hline
$pp\tau^-$&$1(\frac{1}{2}^+)$& $-17078$& B &$[0,{1\over 2}]_{1\over 2}$&$0.22$&$0.16$&$\cdots$
\\ \cline{3-8}
&& $-4807-8.45i$& R &$[0,{1\over 2}]_{1\over 2}$&$0.57$&$0.35$&$\cdots$
\\ \cline{3-8}
&& $-2760-23.72i$& R &$[0,{1\over 2}]_{1\over 2}$&$0.81$&$0.56$&$\cdots$
\\ \cline{3-8}
&&$-2301-1.97i$& R &$[0,{1\over 2}]_{1\over 2}$&$1.17$&$0.71$&$\cdots$
\\ \cline{3-8}
&&$-1880-1.83i$& R &$[0,{1\over 2}]_{1\over 2}$&$1.77$&$1.11$&$\cdots$
\\ \hline\hline
\end{tabular*}
\label{tab:ppl}
\end{center}
\end{table*}

We obtain a bound state with spin-parity $J^P=1/2^+$. Its binding energy is $-2780$ eV, consistent with previous study~\cite{liverts2013three}. The spin configuration of this state is $[s_{12},s_3]_{s}=[0,1/2]_{1/2}$ in the absence of the spin-spin interaction, which indicates that the spins of the two identical protons are anti-aligned. Similar to the hydrogen molecular ion, the bound state in $pp\mu^-$ system also exhibits a symmetric spatial wave function. This result is expected, as an antisymmetric spatial wave function would lead to higher energy levels. The finding is also consistent with the symmetry of spatial wave functions in bound states of $l^+l^+l^{\prime -}$ systems~\cite{Ma:2025rvj}. The rms radius between proton and muon is slightly smaller than the radius between the two protons, indicating a relatively uniform spatial distribution of the three particles within the bound state. In Table~\ref{tab:mass}, we have already obtained the rms radius of the hydrogen-like atom $p\mu^-$. The addition of a second proton in the hydrogen-like molecular ion leads to an additional attractive interaction for the muon, which pulls the muon away from the original proton and slightly increases the $p$–$\mu^-$ separation compared to that in the isolated $p\mu^-$ atom.

We also identify several quasi-bound states with $J^P=1/2^+$. Below the $[p\mu^-](2S)p$ threshold, two resonant states with vanishing decay widths are observed, as shown in Fig.~\ref{fig:ppmu}. Since we ignore the weak interaction in the present work, they can only decay into the $[p\mu^-](1S)p$ channel. Their zero widths are likely due to the orthogonality between the initial and final spatial wave functions, rather than the omission of the $P$-wave decay channel, since the energy of the $p\mu^-(1P)$ state is degenerate with that of the $p\mu^-(2S)$ state. Notably, the binding energy of the first quasi-bound state in our calculation is higher than that reported in previous studies~\cite{liverts2013three,kilic2004coulombic}. By refining the Gaussian basis parameters and increasing the basis size, more accurate energies can be obtained. Between the $[p\mu^-](2S)p$ and $[p\mu^-](3S)p$ thresholds, we observe a dense spectrum of quasi-bound states. As the precision of the calculation improves, the number of additional quasi-bound sates increases correspondingly. In Table~\ref{tab:ppl}, we only provide the energy levels, spin configurations, and rms radii information for the six quasi-bound states. All of these share the same spin configuration: $[s_{12},s_3]_{s}=[0,1/2]_{1/2}$, indicating anti-aligned proton spins. In terms of spatial distribution,
the distance between the two protons exceeds the proton-muon distance but is smaller than twice this value. Although the proton is heavier than the muon, this does not imply that the $pp$ cluster is more compact than the $p\mu^-$ pair. On the contrary, due to the repulsive Coulomb interaction between the two protons and the attractive force between the proton and the muon, the protons are pushed apart from each other while being pulled closer to the muon.

\begin{figure}[hbtp]
\begin{center}
\scalebox{0.35}{\includegraphics{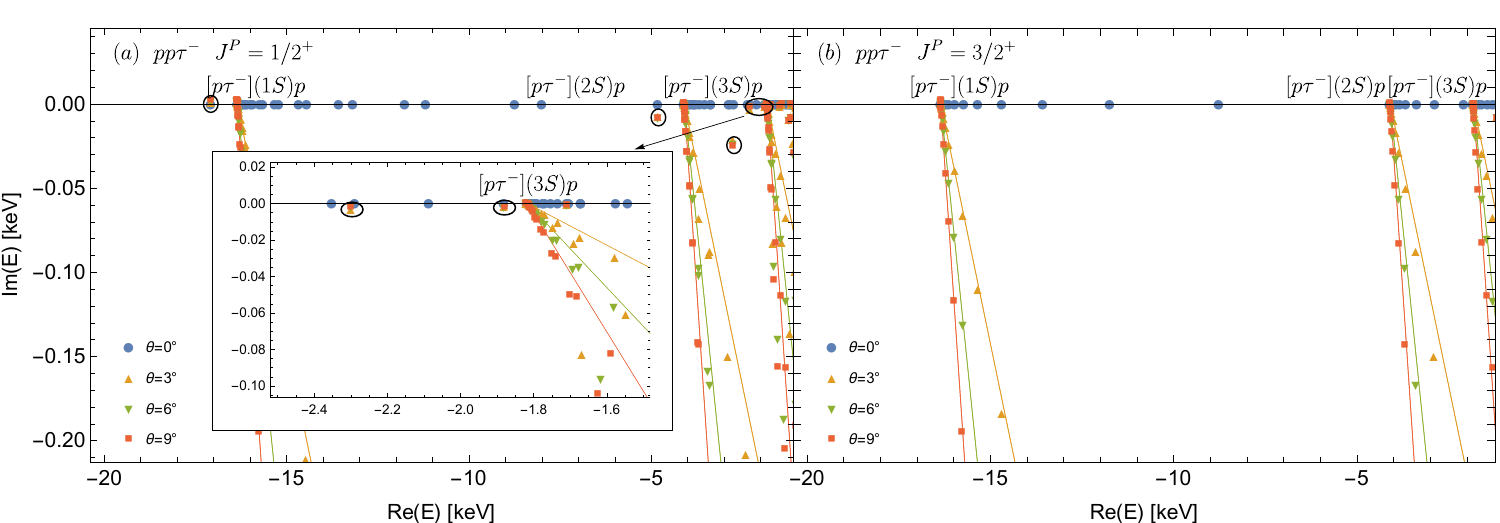}}
\end{center}
\caption{The complex eigenenergies of the $pp\tau^-$ states with varying $\theta$ in the CSM. }
\label{fig:pptau}
\end{figure}

In the $S$-wave $pp\tau^-$ system, we identify one bound state and four quasi-bound states with quantum number $J^P=1/2^+$ as shown in Fig.~\ref{fig:pptau}. Their complex energies, spin configuration, and rms radii are summarized in Table~\ref{tab:ppl}. No bound state or quasi-bound state is observed below the $[p\tau^-](3S)p$ threshold in the sector with quantum number $J^P=3/2^+$.

Compared to the $pp\mu^-$ system, the binding energy of the $pp\tau^-$ bound state is deeper. Since the mass of $\tau^-$ lepton is greater than that of the $\mu^-$, the rms radii between $p$ and $\tau^-$ is significantly smaller than that between the $p$ and $\mu^-$, resulting in stronger Coulomb interactions within the system. Additionally, the distance between two protons is slightly larger than the proton–$\tau^-$ distance, a spatial feature also observed in the bound state of the $pp\mu^-$ system. For the rest of four quasi-bound below the $[p\tau^-](3S)p$ threshold, nonzero decay widths are obtained, although these are underestimated due to the omission of the $\tau^-$ weak decay process in our framework. All of these states exhibit anti-aligned spin configurations between the two protons,i.e., $[s_{12},s_3]_{s}=[0,1/2]_{1/2}$, which implies that the spatial wave function between the two identical protons must be symmetric.

\subsection{$\mu^-\mu^-p$ and $\tau^-\tau^-p$}

\begin{table*}[hbtp]
\begin{center}
\renewcommand{\arraystretch}{1.5}
\caption{The complex eigenenergies $\Delta E-i{\Gamma/2}$, spin components, and rms radii between different particles of the three-body hydrogen-like  $l^-l^- p$ systems. In the last column, we list the binding energies of the corresponding states extracted from other works for comparison.}
\begin{tabular*}{\hsize}{@{}@{\extracolsep{\fill}}cccccccc@{}}
\hline\hline
system ~~~&$I(J^P)$&~~~$\Delta E-i\Gamma/2(\rm eV) $ ~~~&~~type~~&~~spin configuration~~&~~~$r_{l^- l^-}^{\rm{rms}}$(pm)~~~&~~~$r_{l^- p}^{\rm{rms}}$(pm)~~&$\Delta E-i\Gamma/2(\rm eV) $
\\  \hline
$\mu^-\mu^- p$&$\frac{1}{2}(\frac{1}{2}^+)$& $-2655$& B &$[0,{1\over 2}]_{1\over 2}$&$1.46$&$1.02$&$-2655$~\cite{liverts2013three,frolov1993algebraic,ho1979autoionisation,ho1981complex}
\\ \cline{3-8}
&& $-749-3.60i$& R &$[0,{1\over 2}]_{1\over 2}$&$3.95$&$2.37$&$-753-3.57i$~\cite{liverts2013three,frolov1993algebraic,ho1979autoionisation,ho1981complex}
\\ \cline{3-8}
&& $-511-1.22i$& R &$[0,{1\over 2}]_{1\over 2}$&$6.09$&$4.81$&$\cdots$
\\ \cline{3-8}
&&$-498-0.31i$& R &$[1,{1\over 2}]_{1\over 2}$&$7.29$&$5.73$&$\cdots$
\\ \cline{3-8}
&&$-468-0.76i$& R &$[0,{1\over 2}]_{1\over 2}$&$8.27$&$6.38$&$\cdots$
\\ \cline{3-8}
&&$-448-0.78i$& R &$[1,{1\over 2}]_{1\over 2}$&$9.84$&$7.58$&$\cdots$
\\ \cline{3-8}
&&$-427-0.27i$& R &$[0,{1\over 2}]_{1\over 2}$&$10.74$&$8.22$&$\cdots$
\\ \cline{3-8}
&&$-406-1.07i$& R &$[1,{1\over 2}]_{1\over 2}$&$12.31$&$9.42$&$\cdots$
\\ \cline{3-8}
&&$-390-0.15i$& R &$[0,{1\over 2}]_{1\over 2}$&$13.22$&$10.09$&$\cdots$
\\ \cline{3-8}
&&$-371-1.28i$& R &$[1,{1\over 2}]_{1\over 2}$&$14.81$&$11.31$&$\cdots$
\\ \cline{3-8}
&&$-358-0.14i$& R &$[0,{1\over 2}]_{1\over 2}$&$15.62$&$11.90$&$\cdots$
\\ \cline{2-8}
&$\frac{1}{2}(\frac{3}{2}^+)$& $-498-0.31i$& R &$[1,{1\over 2}]_{3\over 2}$&$7.29$&$5.73$&$\cdots$
\\ \cline{3-8}
&& $-448-0.78i$& R &$[1,{1\over 2}]_{3\over 2}$&$9.84$&$7.58$&$\cdots$
\\ \cline{3-8}
&& $-406-1.07i$& R &$[1,{1\over 2}]_{3\over 2}$&$12.31$&$9.42$&$\cdots$
\\ \cline{3-8}
&& $-371-1.28i$& R &$[1,{1\over 2}]_{3\over 2}$&$14.81$&$11.31$&$\cdots$
\\ \cline{3-8}
&& $-340-1.24i$& R &$[1,{1\over 2}]_{3\over 2}$&$17.46$&$13.31$&$\cdots$
\\ \hline
$\tau^-\tau^- p$&$\frac{1}{2}(\frac{1}{2}^+)$& $-17272$& B &$[0,{1\over 2}]_{1\over 2}$&$0.19$&$0.14$&$\cdots$
\\ \cline{3-8}
&& $-4902-0.23$& R &$[0,{1\over 2}]_{1\over 2}$&$0.52$&$0.33$&$\cdots$
\\ \cline{3-8}
&& $-2853-10.07$& R &$[0,{1\over 2}]_{1\over 2}$&$0.77$&$0.56$&$\cdots$
\\ \cline{3-8}
&&$-2478-25.35$& R &$[0,{1\over 2}]_{1\over 2}$&$0.96$&$0.62$&$\cdots$
\\ \cline{3-8}
&&$-2116-12.90$& R &$[0,{1\over 2}]_{1\over 2}$&$1.30$&$0.79$&$\cdots$
\\ \hline\hline
\end{tabular*}
\label{tab:llp}
\end{center}
\end{table*}

\begin{figure}[hbtp]
\begin{center}
\scalebox{0.35}{\includegraphics{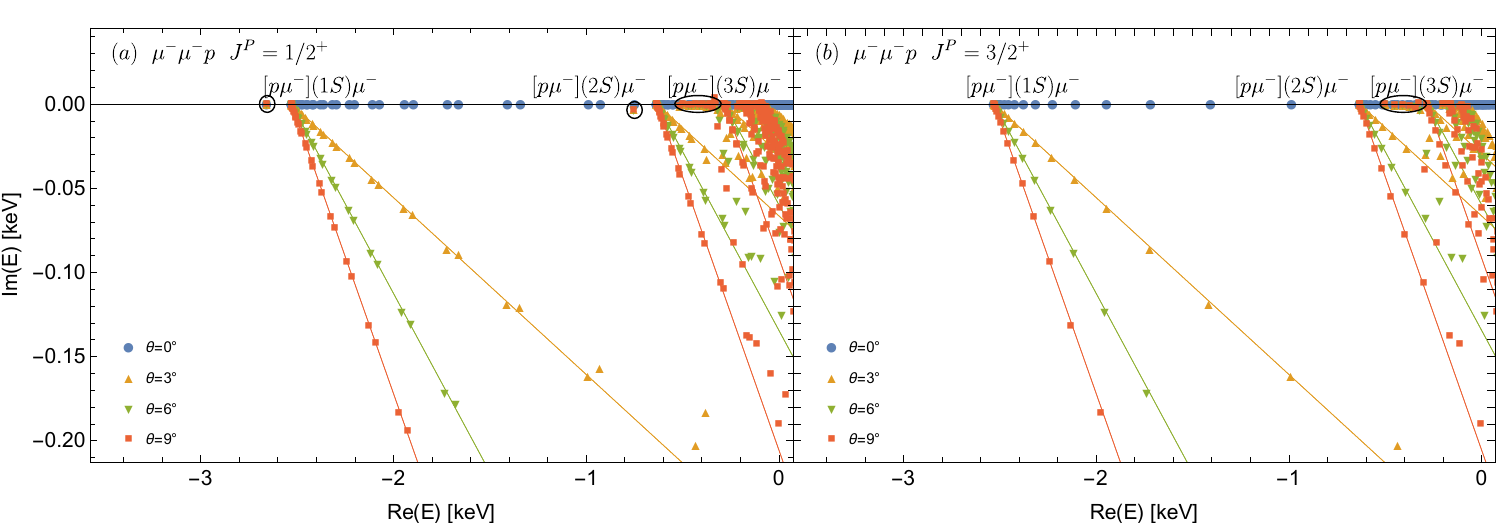}}
\end{center}
\caption{The complex eigenenergies of the $\mu^-\mu^-p$ states with varying $\theta$ in the CSM.}
\label{fig:mumup}
\end{figure}

\begin{figure}[hbtp]
\begin{center}
\scalebox{0.35}{\includegraphics{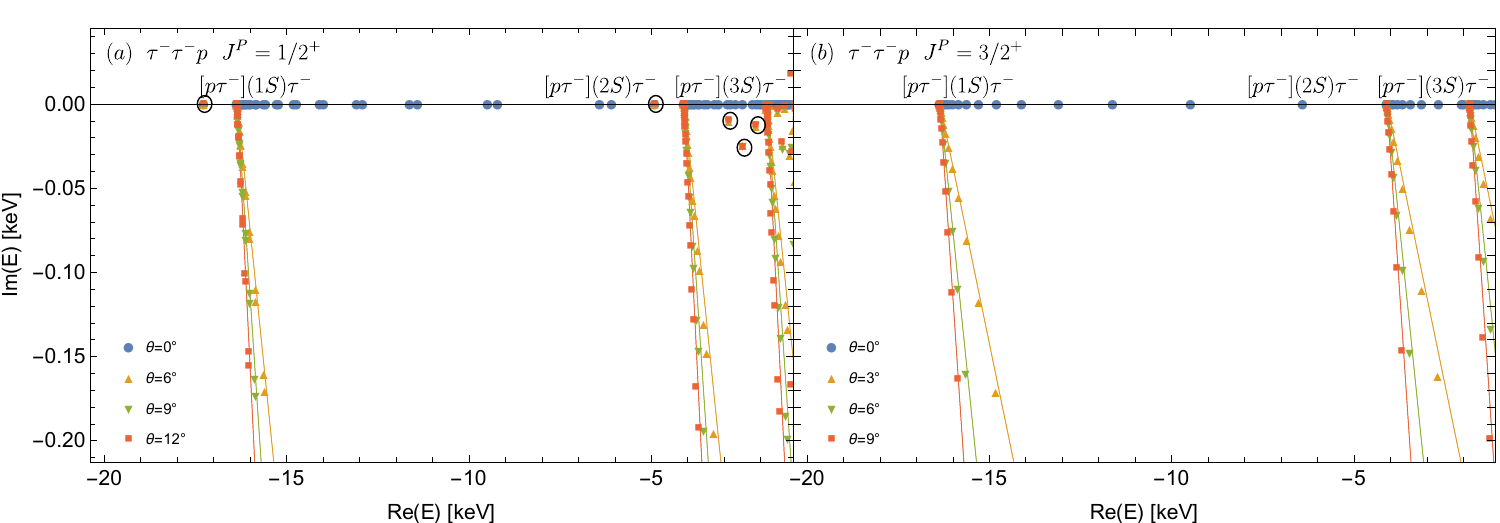}}
\end{center}
\caption{The complex eigenenergies of the $\tau^-\tau^-p$ states with varying $\theta$ in the CSM. }
\label{fig:tautaup}
\end{figure}

We investigate the complex eigenenergies of $S$-wave $l^-l^-p$ systems within the CSM framework. For the $\mu^-\mu^-p$ system, we obtain one bound state and several quasi-bound states with quantum numbers $J^P=1/2^+$, and $3/2^+$. In the $\tau^-\tau^-p$ system, we also identify a bound state and four quasi-bound states with $J^P=1/2^+$. Their binding energies $\Delta E$ range from $-17.3$~keV to $-3.4$~keV relative to the three body threshold. The results for $\mu^-\mu^-p$ and $\tau^-\tau^-p$ systems are shown in Figs.~\ref{fig:mumup}, \ref{fig:tautaup}. The complex eigenenergies, spin components and rms radii of different particles are listed in Table~\ref{tab:llp}.

In the $\mu^-\mu^-p$ system, the bound state and the first quasi-bound state lie below the $[p\mu^-](2S)\mu^-$ threshold and share the spin configuration $[s_{12},s_3]_{s}=[0,1/2]_{1/2}$. The binding energy of the bound state is $-2655$~eV, which is approximately consistent with the result reported in Refs.~\cite{liverts2013three,frolov1993algebraic,ho1979autoionisation,ho1981complex}. Since the proton and muons carry opposite charges, the proton is attracted by both $\mu^-$ particles. As a result, the rms distance between the two muons is larger than that between the proton and a muon. The higher quasi-bound states are located on the second Riemann sheet of the $[p\mu^-](2S)\mu^-$ channel and the first Riemann sheet of the $[p\mu^-](3S)\mu^-$ channel. We analyze the complex energies, spin configurations, and rms radii of the ten quasi-bound states. Several states exhibit aligned spins for the two muons. Owing to the Pauli principle, such spin-aligned configurations must have antisymmetric spatial wave functions.

In addition, we identify quasi-bound states in the $\mu^-\mu^-p$ system with quantum numbers $J^P=3/2^+$. They are distributed between the $[p\mu^-](2S)\mu^-$ and $[p\mu^-](3S)\mu^-$ thresholds, and all share the spin configuration $[s_{12},s_3]_{s}=[1,1/2]_{3/2}$. In contrast to other QED systems with with $J^P=3/2^+$, such as $l^+l^+l^{(\prime)-}$~\cite{Ma:2025rvj} and $ppl^-$, only the $\mu^-\mu^-p$ system exhibits quasi-bound states.

The results of $\tau^-\tau^-p$ system are presented in Fig.~\ref{fig:tautaup}. We identify a bound state and four quasi-bound states with $J^P=1/2^-$ below the $[p\tau^-](3S)\tau^-$ threshold. Their complex energies, spin configurations, and rms radii are summarized in Table~\ref{tab:llp}. All obtained states have spin configurations $[s_{12},s_3]_{s}=[0,1/2]_{1/2}$. The binding energy of the $\tau^-\tau^-p$ bound state is approximately six times larger than that of the $\mu^-\mu^-p$ system. This deeper binding is expected, as the larger mass and lower kinetic energy of the $\tau^-$ leptons result in smaller $\tau^-\tau^-$ and $p\tau^-$ distances and hence stronger binding. Furthermore, a quasi-bound state appears below the $[p\tau^-](2S)\tau^-$ threshold, which probably  corresponds to the tauonic analogue of the resonance below the $2S$ threshold in the $\mu^-\mu^-p$ system.

\subsection{$p \mu^-\tau^-$}

\begin{figure}[hbtp]
\begin{center}
\scalebox{0.35}{\includegraphics{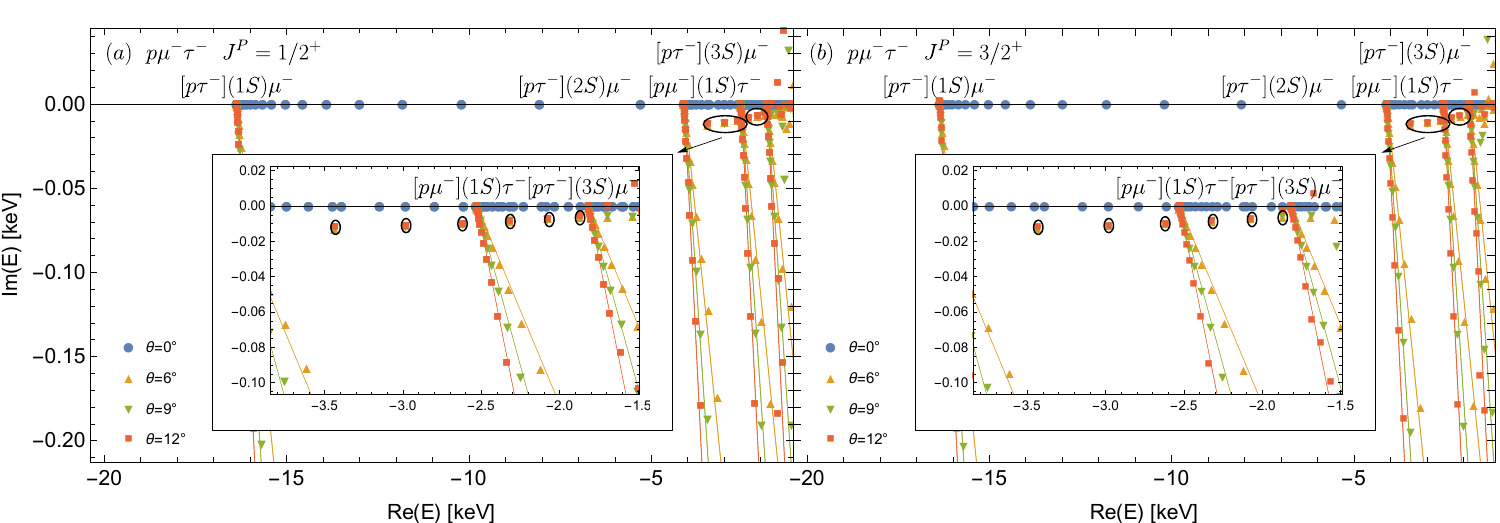}}
\end{center}
\caption{The complex eigenenergies of the $p\mu^-\tau^-$ states with varying $\theta$ in the CSM.  }
\label{fig:pmutau}
\end{figure}

We calculate the complex energies of the $S$-wave $p\mu^-\tau^-$ system within the CSM framework and identify several quasi-bound states below $[p\tau^-](3S)$ threshold with quantum numbers $J^P=1/2^+$ and $3/2^+$. The results for the $p\mu^-\tau^-$ system are shown in Fig.~\ref{fig:pmutau}, and the complex eigenenergies, spin configurations, and rms radii are listed in Table~\ref{tab:pmutau}. Unlike the $ppl^-$ and $pl^-l^-$ systems, these quasi-bound states contain no identical particles and therefore are not subject to the Pauli principle. In the nonrelativistic Hamiltonian, we only use the Coulomb potential and neglect the spin-spin interaction. As a result, states with different spin configurations, such as $[s_{12},s_3]_{s}=[0,1/2]_{1/2}$, $[s_{12},s_3]_{s}=[1,1/2]_{1/2}$, and $[s_{12},s_3]_{s}=[1,1/2]_{3/2}$, exhibit degenerate complex eigenenergies. All quasi-bound states share a notable spatial characteristic: the rms radii $r_{p\mu^-}$ and $r_{\mu^-\tau^-}$ are of comparable magnitude, while $r_{p\tau^-}$ is significantly smaller. This pattern suggests a quasi-isosceles triangle spatial arrangement of the three particles, with the proton positioned much closer to the $\tau^-$ lepton.

\begin{table*}[hbtp]
\begin{center}
\renewcommand{\arraystretch}{1.5}
\caption{The complex eigenenergies $\Delta E-i{\Gamma/2}$, spin components, and rms radii between different particles of the three-body hydrogen-like $p\mu^- \tau^-$ systems.}
\begin{tabular*}{\hsize}{@{}@{\extracolsep{\fill}}cccccccc@{}}
\hline\hline
system ~~~&$I(J^P)$&~~~$\Delta E-i\Gamma/2(\rm eV) $ ~~~&~~type~~&~~~spin configuration~~~&~~~$r_{p \mu^-}^{\rm{rms}}$(pm)~~~&~~~$r_{p \tau^-}^{\rm{rms}}$(pm)~~~&~~~$r_{\mu^- \tau^-}^{\rm{rms}}$(pm)~~
\\  \hline
$p\mu^-\tau^- $&$\frac{1}{2}(\frac{1}{2}^+)/(\frac{3}{2}^+)$& $-3432-12.30i$& R &$[0,{1\over 2}]_{1\over 2}/[1,{1\over 2}]_{1\over 2}/[1,{1\over 2}]_{3\over 2}$&$1.91$&$0.27$&$1.92$
\\ \cline{3-8}
&& $-2978-11.34i$& R &$[0,{1\over 2}]_{1\over 2}/[1,{1\over 2}]_{1\over 2}/[1,{1\over 2}]_{3\over 2}$&$2.91$&$0.32$&$2.98$
\\ \cline{3-8}
&& $2614-10.17i$& R &$[0,{1\over 2}]_{1\over 2}/[1,{1\over 2}]_{1\over 2}/[1,{1\over 2}]_{3\over 2}$&$3.83$&$0.36$&$3.94$
\\ \cline{3-8}
&&$-2319-8.30i$& R &$[0,{1\over 2}]_{1\over 2}/[1,{1\over 2}]_{1\over 2}/[1,{1\over 2}]_{3\over 2}$&$4.73$&$0.46$&$4.89$
\\ \cline{3-8}
&&$-2074-7.37i$& R &$[0,{1\over 2}]_{1\over 2}/[1,{1\over 2}]_{1\over 2}/[1,{1\over 2}]_{3\over 2}$&$5.70$&$0.48$&$5.91$
\\ \cline{3-8}
&&$-1867-6.41i$& R &$[0,{1\over 2}]_{1\over 2}/[1,{1\over 2}]_{1\over 2}/[1,{1\over 2}]_{3\over 2}$&$6.71$&$0.55$&$6.98$
\\ \hline\hline
\end{tabular*}
\label{tab:pmutau}
\end{center}
\end{table*}

\section{Four-body hydrogen-like systems}
\label{sec:results-1}

\begin{figure*}[hbtp]
\begin{center}
\scalebox{0.6}{\includegraphics{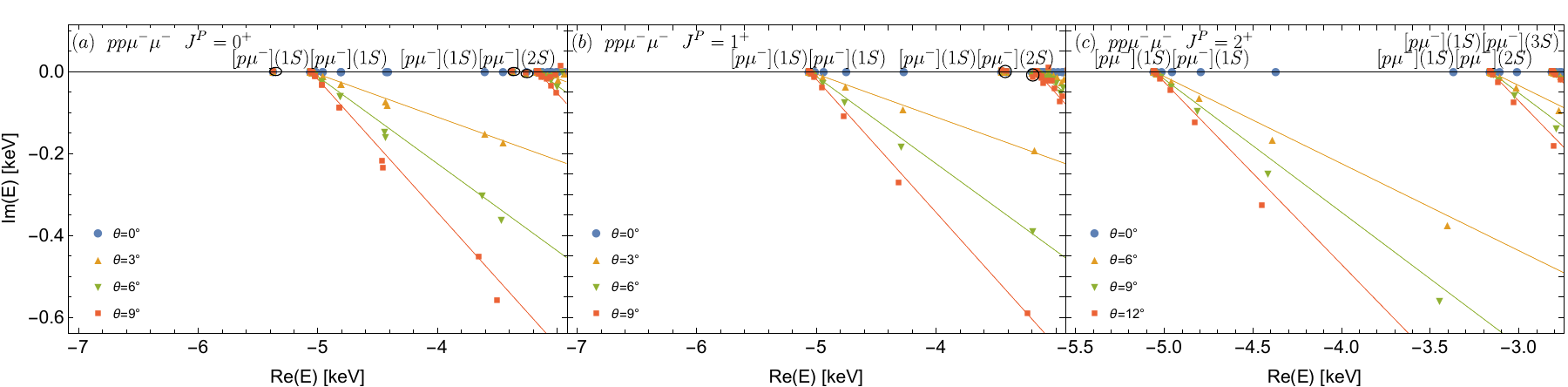}}
\end{center}
\caption{The complex eigenenergies of the $pp\mu^-\mu^-$ states with varying $\theta$ in the CSM.}
\label{fig:ppmumu}
\end{figure*}

\begin{figure*}[hbtp]
\begin{center}
\scalebox{0.6}{\includegraphics{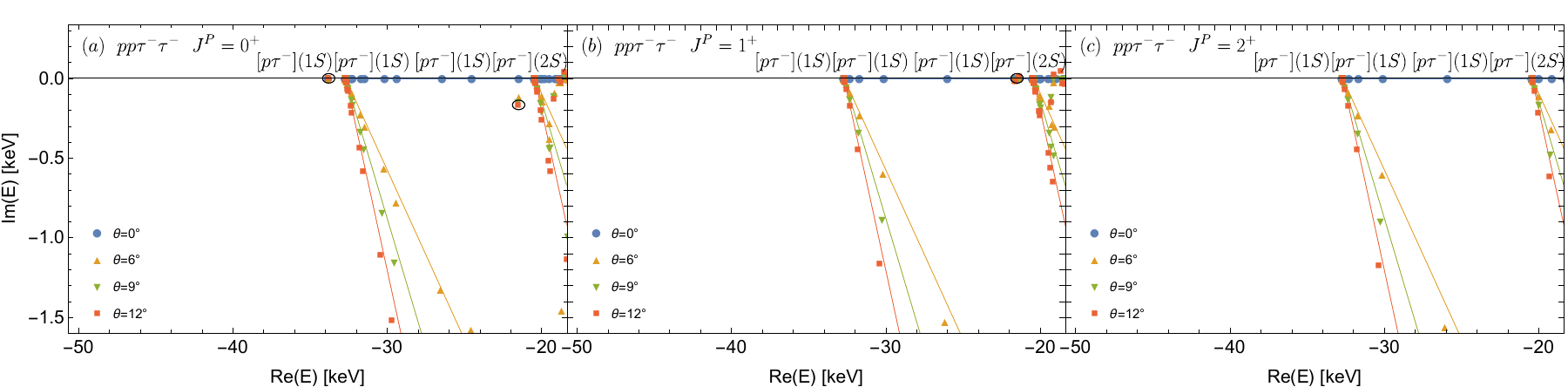}}
\end{center}
\caption{The complex eigenenergies of the $pp\tau^-\tau^-$ states with varying $\theta$ in the CSM.}
\label{fig:pptautau}
\end{figure*}

We investigate the complex eigenenergies of $S$-wave $ppl^-l^-$ systems within the CSM framework. The results for $pp\mu^-\mu^-$ and $pp\tau^-\tau^-$ systems with quantum numbers $J^P=0^+$, $1^+$, and $2^+$ are shown in Figs.~\ref{fig:ppmumu}, \ref{fig:pptautau}, respectively. In the $pp\mu^-\mu^-$ system, we obtain a bound state and five quasi-bound states below the $[p\mu^-](1S)[p\mu^-](2S)$ threshold. For the $pp\tau^-\tau^-$ system, we find a bound state and three quasi-bound states below the $[p\tau^-](1S)[p\tau^-](2S)$ threshold. Their complex eigenenergies, spin components, and rms radii are listed in Tables~\ref{tab:ppmumu}, \ref{tab:pptautau}.
 
\subsection{$pp\mu^-\mu^-$}

For the four-body system $pp\mu^-\mu^-$, which contains two pairs of identical particles, we systematically analyze their spatial configurations. For computational convenience, we focus only on the hydrogen-like atomic pair structure and two representative $K$-type configurations, $[[pl^-]p]l^-$ and $[[pl^-]l^-]p$, as well as the additional configurations generated through particle exchange symmetry. We construct ten spatial configurations, each described using different sets of Jacobi coordinates $r_\alpha$, $\lambda_\alpha$, and $\rho_\alpha$, as illustrated in Fig~\ref{fig:Jacobi-four}. Among these, two configurations are di-$pl^-$ type, while the remaining eight correspond to various $K$-type structures, labeled in panels (a)–(j). In particular, exchanging identical particles transforms the configuration in panel (a) into that in panel (b), and those in panels (c) and (d) can be transformed into six additional $K$-type clusters through similar exchanges. In contrast, within tetraquark systems, given that quarks(antiquark) possess color degrees of freedom, the color confinement prohibits the existence of free quarks. Consequently, in our previous work~\cite{Chen:2023syh,Ma:2024vsi,Wu:2024euj,Wu:2024hrv,Wu:2024zbx,Yang:2025wqo}, we neglected the $K$-type configuration. However, $K$-type arrangements may become necessary in four-body QED systems, where electromagnetic interaction between one particle and the remaining three-particle cluster can occur over long ranges. 

We obtain a bound state and some quasi-bound states with $J^P=0^+$ and $1^+$ below the $[p\mu^-](1S)[p\mu^-](2S)$ threshold. In the $J^P=0^+$ sector, the binding energy of the bound state is $5366$~keV relative to the four-body threshold. It exhibits a spin configuration of $[s_{12},s_{34}]_{s}=[0,0]_{0}$, indicating that the spins of both the $pp$ and $\mu^-\mu^-$
pairs are antiparallel. This leads to a symmetric spatial wave function. Analogous to the hydrogen molecule, the two muons are shared by the two protons, which results in a strong covalent bond. The rms radius $r_{p\mu^-}$ in this bound state is larger than that in a hydrongen-like $p\mu^-(1S)$ atom. The distance between the two protons is slightly larger than $r_{p\mu^-}$, similar to the behavior observed in the $pp\mu^-$ system, while the separation between two muons is relatively large. Since the proton mass is approximately nine times greater than that of the muon, the kinetic energy of the protons is significantly lower than that of muons. The combined effect of this mass difference and the Coulomb interaction leads to the observed spatial distribution of this state. For the two higher quasi-bound states, their spin components remain $[s_{12},s_{34}]_{s}=[0,0]_{0}$. The rms radii $r_{pp}$, $r_{p\mu^-}$, and $r_{\mu^-\mu^-}$ all exceed $1$~pm, with $r_{p\mu^-}$ being slightly larger than $r_{pp}$.

In the $J^P=1^+$ system, we identify three quasi-bound states below the $[p\mu^-](1S)[p\mu^-](2S)$ threshold, but no bound state is found. These states exhibit spin configurations of either $[s_{12},s_{34}]_{s}=[1,0]_{1}$ or $[s_{12},s_{34}]_{s}=[0,1]_{1}$, indicating that one pair of identical particles has aligned spins, and the associated spatial wave function must be antisymmetric. Such configurations do not favor the formation of a bound state in the $J^P=1^+$ system but can support the formation of some quasi-bound states. In the $J^P=2^+$ sector, we find neither bound states nor quasi-bound states in the $pp\mu^-\mu^-$ system. Consistent with the $\mu^+\mu^+ e^-e^-$ system with $J^P=2^+$~\cite{Ma:2025rvj}, no bound states or resonant states are found in the present configuration.

Although studies of the charmonium-like state $T_{cc}(3875)$~\cite{LHCb:2021vvq,LHCb:2021auc} offer valuable methodological insights for exploring the $pp\mu^-\mu^-$ system, the bound state in the $pp\mu^-\mu^-$ system differs significantly from those in doubly heavy tetraquark systems. The $T_{cc}(3875)$, composed of two heavy quarks and two light quarks, is a shallow bound state with quantum numbers $I(J^P)=0(1^+)$~\cite{Wu:2024zbx}. In our calculation, the $pp\mu^-\mu^-$ system, which similarly contains two heavy and two light particles, is found to support a bound state only in the $J^P=0^+$ channel. This discrepancy is attributable to the complex chromoelectric and chromomagnetic interactions in the QCD-based systems, in contrast to the purely Coulombic interactions governing QED systems.

 
In this system, the $K$-type Jacobi spatial structure has a significant impact on energies of bound and quasi-bound states. We find that including $K$-type configurations lowers the binding energy of the bound state by approximately $36$~eV compared to the calculation that excludes them. Additionally, an extra quasi-bound pole appears below the $[p\mu^-](1S)[p\mu^-](2S)$ threshold, and the lowest quasi-bound state is stabilized by $60$~eV relative to the non-$K$-type calculation. These results highlight the critical role of $K$-type Jacobi coordinates in QED-based few-body systems. Physically, expanding the configuration space of the wave function by incorporating more complete spatial degrees of freedom leads to lower energies.


\subsection{$pp\tau^-\tau^-$}

In the $pp\tau^-\tau^-$ system, we identify one bound state and three quasi-bound states with $J^P=0^+$ and $1^+$, all of which are located below the $[p\tau^-](1S)[p\tau^-](2S)$ threshold. Similar to the $pp\mu^-\mu^-$ system with $0^+$, the bound state has a spin configuration of $[s_{12},s_{34}]_{s}=[0,0]_{0}$ and can be regarded as the tauonic analogue of the bound state in the $pp\mu^-\mu^-$ system. Due to the larger mass of the $\tau^-$ lepton, the binding energy in the $pp\tau^-\tau^-$ is significantly deeper than that in the muonic case. For the quasi-bound state in the $J^P=0^+$ sector, the spin configuration is also $[s_{12},s_{34}]_{s}=[0,0]_{0}$, and it exhibits a relatively uniform spatial distribution, with comparable rms radii $r_{p\tau^-}$, $r_{pp}$, and $r_{\tau^-\tau^-}$.

For the $J^P=1^+$ sector, we identify two quasi-bound states below the $[p\tau^-](1S)[p\tau^-](2S)$ threshold. Their spin configurations are $[s_{12},s_{34}]_{s}=[0,1]_{1}$ and $[1,0]_{1}$, respectively. As in the $pp\mu^-\mu^-$ or $\mu^+\mu^+ e^-e^-$ systems, bound states are not supported in this channel, but quasi-bound states may exist. The different particle masses in these QED systems likely result in the different number of observed quasi-bound states. Notably, the lowest quasi-bound state with $J^P=1^+$ shows distinctive spatial characteristics, where the rms radii $r_{pp}< r_{p\tau^-}< r_{\tau^-\tau^-} $. Given that the mass of the $\tau^-$ is greater than that of the proton, the two $\tau^-$ particles can be regarded as nearly stationary charges, with the two protons shared between them. Contrary to expectations based solely on mass-scaling arguments, the Coulomb repulsion may enforce $r_{\tau^-\tau^-}>r_{p\tau^-}$. The separation of the protons would weaken the attractive $\tau^-$-$p$ interaction, thereby driving them further apart. Consequently, this geometric reorganization results in a more diffuse spatial structure and leads to an higher quasi-bound state energy. For the $J^P=2^+$ sector, neither bound or resonant states are found.

Analogous to the $pp\mu^-\mu^-$ system, $K$-type spatial configurations are also essential for accurately describing bound and quasi-bound states in the hydrogen-like $pp\tau^-\tau^-$ molecular system. In addition to the hydrogen-like atomic pair structures ($[p\tau^-][p\tau^-]$), our computational framework incorporates two $K$-type configurations that involve hydrogen-like atomic threshold $[[p\tau^-]\tau^-]p$ and $[[p\tau^-]p]\tau^-$. The full $K$-type bases are truncated to maintain computational tractability. Nevertheless, the inclusion of these specific configurations induces significant energy shifts in both bound and resonant states. For the $J^P=0^+$ system, the inclusion of the $K$-type configurations lowers the bound state energy by approximately $40$~eV and reduces the quasi-bound state energy by about $50$~eV. In the $J^P=1^+$ sector, incorporating these $K$-structures results in the appearance of two additional resonant poles below the $[p\tau^-](1S)[p\tau^-](2S)$ threshold.

\begin{table*}[hbtp]
\begin{center}
\renewcommand{\arraystretch}{1.5}
\caption{The complex eigenenergies $\Delta E-i{\Gamma/2}$, spin components, and rms radii between different particles of the four-body hydrogen-like $pp \mu^-\mu^-$ systems. The binding energies shown in the third column are obtained by including the $K$-type spatial configurations, whereas the values in the last column are calculated without the $K$-type configurations.}
\begin{tabular*}{\hsize}{@{}@{\extracolsep{\fill}}ccccccccc@{}}
\hline\hline
system ~~~&$I(J^P)$&~~~$\Delta E-i\Gamma/2(\rm eV) $ ~~~&~~type~~&~~~spin configuration~~~&$r_{p \mu^-}^{\rm{rms}}$(pm)&~~$r_{p p}^{\rm{rms}}$(pm)~~&~~$r_{\mu^- \mu^-}^{\rm{rms}}$(pm)~~& $\Delta E-i\Gamma/2(\rm eV) $
\\  \hline
$pp\mu^-\mu^-$&$1(0^+)$& $-5366$& B &$[0,0]_0$&$0.63$&$0.66$&$0.83$& $-5330$
\\ \cline{3-9}
&& $-3370-0.16i$& R &$[0,0]_0$&$1.44$&$1.23$&$1.77$&$-3311-2.19i$
\\ \cline{3-9}
&& $-3259-4.30i$& R &$[0,0]_0$&$1.53$&$1.44$&$1.84$ &$\cdots$
\\ \cline{2-9}
&$1(1^+)$&$-3449$& R &$[1,0]_{1}$&$1.14$&$1.22$&$1.20$&$-3316$ 
\\ \cline{3-9}
&&$-3406$& R &$[0,1]_{1}$&$1.41$&$0.83$&$1.93$ &$\cdots$
\\ \cline{3-9}
&&$-3194-7.30i$& R &$[1,0]_{1}$&$1.66$&$2.03$&$1.83$ &$\cdots$
\\ \hline\hline
\end{tabular*}
\label{tab:ppmumu}
\end{center}
\end{table*}

\begin{table*}[hbtp]
\begin{center}
\renewcommand{\arraystretch}{1.5}
\caption{The complex eigenenergies $\Delta E-i{\Gamma/2}$, spin components, and rms radii between different particles of the four-body hydrogen-like $pp \tau^-\tau^-$ systems. In the third column, the binding energies are obtained by including the $K$-type spatial configurations, while the complex eigenenergies in the last column are calculated without the $K$-type configurations.}
\begin{tabular*}{\hsize}{@{}@{\extracolsep{\fill}}ccccccccc@{}}
\hline\hline
system ~~~&$I(J^P)$&~~~$\Delta E-i\Gamma/2(\rm keV) $ ~~~&~type~&~~spin configuration~~&$r_{p \tau^-}^{\rm{rms}}$(pm)&~~$r_{p p}^{\rm{rms}}$(pm)~~&~~$r_{\tau^- \tau^-}^{\rm{rms}}$(pm)~~&$\Delta E-i\Gamma/2(\rm keV) $
\\  \hline
$pp\tau^-\tau^-$&$1(0^+)$& $-33.8$& B &$[0,0]_0$&$0.12$&$0.15$&$0.14$&$-33.4$
\\ \cline{3-9}
&& $-21.5-0.2i$& R &$[0,0]_0$&$0.24$&$0.26$&$0.26$&$-21.0-0.1i$
\\ \cline{2-9}
&$1(1^+)$&$-21.6-0.01i$& R &$[0,1]_{1}$&$0.21$&$0.17$&$0.27$&$\cdots$
\\ \cline{3-9}
&&$-21.3$& R &$[1,0]_{1}$&$0.23$&$0.30$&$0.17$&$\cdots$
\\ \hline\hline
\end{tabular*}
\label{tab:pptautau}
\end{center}
\end{table*}

\section{Summary}
\label{sec:summary}
In this work we investigate the mass spectrum of $S$-wave three-body hydrogen-like negative ion systems $\mu^-\mu^- p$, $\tau^-\tau^- p$, $p\mu^-\tau^- $, and three-body hydrogen-like molecular ion systems $pp\mu^-$, $pp\tau^-$, as well as four-body hydrogen-like molecular systems $pp\mu^-\mu^-$, $pp\tau^-\tau^-$ using the QED Coulomb potential. We consider a complete set of spin basis functions and different sets of Jacobi coordinates, with twenty Gaussian basis functions for each coordinate. The complex-scaled  Schr\"{o}dinger equation is solved employing the complex scaling method and the Gaussian expansion method, allowing us to identify possible bound and quasi-bound states.

We obtain six bound states and a series of quasi-bound states in $\mu^-\mu^- p$, $\tau^-\tau^- p$, $p\mu^-\tau^-$, $pp\mu^-$, $pp\tau^-$ systems with quantum numbers $J^P=1/2^+$, $3/2^+$, and $pp\mu^-\mu^-$ and $pp\tau^-\tau^-$ systems with quantum numbers $J^P=0^+$, $1^+$, $2^+$. The binding energies of these states range from $-33.8$~keV to $-340$~eV. Among them, only the $p\mu^-\tau^- $ system has no bound state. The $p\mu^-\tau^-$ system contains no identical particles and we neglect spin-spin interactions. Therefore, the complex eigenenergies for $J^P=1/2^+$ and $3/2^+$ sectors exhibit degeneracy. The quasi-bound states in the $p\mu^-\tau^-$ system lie approximately $0.05-1.6$~keV below the $[p\tau^-](3S)\mu^-$ threshold.
 
There exist quasi-bound states with widths less than $1$ eV. Among these narrow-width states, some three-body system states lie below the $[pl^-](2S)p/[pl^-](2S)l^-$ threshold, and some four-body hydrogen-like molecular states are located below the $[pl^-](1S)[pl^{\prime -}](2S)$ threshold. The suppression of the widths of these quasi-bound states may originate from the significantly suppressed overlap integral between the initial and final state wave function in the electromagnetic transition process. This suppression cannot be attributed to the coupling with the $P$-wave channel(e.g., $[pl^-](1P)$), since $\Delta E_{[pl^-](1P)}=\Delta E_{[pl^-](2S)}$. Conversely, for those quasi-bound states whose energy levels lie above the $2S$ threshold, the narrow widths may also result from their potential decays into the $[pl^-](1P)$ channel, which is not included in the current calculation. A more detailed analysis is required to clarify the suppression mechanisms.

Our calculations reveal that all bound states exhibit exclusive spin configurations of $[0,1/2]_{1/2}$ or $[0,0]_0$. This implies the antiparallel spin alignment among identical particles, enforcing symmetric spatial wavefunctions dominated by $S$-wave components. For the majority of three-body quasi-bound states with $J^P=1/2^+$ and four-body quasi-bound states with $J^P=0^+$, antiparallel spin alignment of identical particles also remains the dominant configuration. However, in the $\mu^-\mu^-p$ with $J^P=1/2^+$, certain quasi-bound states exhibit parallel spin alignment of identical particles. These high-lying states reside above the $[p\mu^-](2S)\mu^-([p\mu^-](1P)\mu^-)$ threshold, suggesting the possible involvement of two $P$-wave orbital excitations, while rigorously conserving total spin and parity. An exception occurs in the $pp\mu^-\mu^-$ and $pp\tau^-\tau^-$ systems with $J^P=1^+$, where one pair of identical particles is anti-aligned while the other pair is aligned.

To describe the spatial configurations of these bound and quasi-bound states, we use rms radii as the indicator to portray the distribution of particles in three-body and four-body systems. For the $S$-wave hydrogen-like systems $pp\mu^-$, $pp\tau^-$, $pp\mu^-\mu^-$, and $pp\tau^-\tau^-$, the bound and quasi-bound states exhibit covalent bond-like characteristic, where one or two leptons are shared by two protons and each proton-lepton pair has identical distance. In the hydrogen-like negative ion systems $\mu^-\mu^-p$ and $\tau^-\tau^-p$, the proton is shared between the two leptons in the bound and quasi-bound states, leading to the same distances in each proton-lepton pair. Notably, the rms radius $r_{p\mu^-}$ is significantly larger than $r_{p\tau^-}$, reflecting the mass difference between the muon and tau. In the $S$-wave $p\mu^-\tau^-$ quasi-bound states, the distance between $p$ and $\mu^-$ is comparable to that between $\mu^-$ and $\tau^-$, and the proton is very close to $\tau^-$.

The $K$-type spatial configurations are essential for accurately describing bound and quasi-bound states in the hydrogen-like molecular systems $pp\mu^-\mu^-$ and $pp\tau^-\tau^-$. In our computational framework, the inclusion of $K$-type configurations significantly alters binding energies of both bound and quasi-bound states. Specifically, for the $pp\mu^-\mu^-$ system with $J^P=0^+$, incorporating $K$-type configurations reduces the binding energy by approximately $36$~eV compared to the calculation that omits them. Additionally, a new quasi-bound state emerges below the $[p\mu^-](1S)[p\mu^-](2S)$ threshold, and the lowest quasi-bound state exhibits a substantial downward energy shift exceeding $60$~eV. Similarly, in the $pp\tau^-\tau^-$ system with $J^P=0^+$, the inclusion of these specific $K$-type configurations lowers the bound-state energy by about $40$~eV and the quasi-bound energy by approximately $50$~eV. In the $J^P=1^+$ sector, the incorporation of these $K$-type structures generates two additional quasi-bound states below the $[p\tau^-](1S)[p\tau^-](2S)$ threshold.

In summary, we use complex scaling and Gaussian expansion methods to investigate three-body and four-body hydrogen-like systems $pp\mu^-$, $pp\tau^-$, $\mu^-\mu^-p$, $\tau^-\tau^-p$, $p\mu^-\tau^-$, $pp\mu^-\mu^-$ and $pp\tau^-\tau^-$. In these systems, the QED Coulomb potential dominates the interparticle interactions, thereby governing their physical properties. This area has stimulated intensified theoretical and experimental interest aimed at uncovering QED effects in such exotic bound and quasi-bound states. Previous studies have primarily focused on the bound states of the $pp\mu^-$ and $\mu^-\mu^-p$ systems, while little attention has been paid to their resonance structures. In this work, we systematically explore both the bound and quasi-bound states across a range of three- and four-body hydrogen-like systems. For the first time, we provide the theoretical estimates of the binding energies of bound and quasi-bound states in $pp\mu^-\mu^-$ and $pp\tau^-\tau^-$, making a significant contribution to the study of exotic few-body Coulomb systems. Looking ahead, future muon colliders and high-intensity muon sources will offer promising opportunities for the possible abundant production of such hydrogen-like molecules and ions. Experimental access to these states could be realized through scattering processes such as: $2\mu^- + \mathrm{H_2} \to \mathrm{H_{2\mu}} + 2e^-$, $\mu^- + \mathrm{H_2} \to \mathrm{H_{\mu e}} + e^-$, and $\mu^- + \mathrm{H_2^+} \to \mathrm{H_{2\mu}^+} + e^-$. These reactions would pave the way for the investigations of novel states like $\mathrm{H_{2\mu}}$, $\mathrm{H_{\mu e}}$, and $\mathrm{H_{2\mu}^+}$, enriching our understanding of QED in multi-particle systems.

\section*{ACKNOWLEDGMENTS}

This project is supported by the China Postdoctoral Science Foundation under Grants No.~2024M750049, and
the National Natural Science Foundation of China under Grants No.~12475137. The computational resources were supported by High-performance Computing Platform of Peking University.

\bibliography{Ref}

\end{document}